\documentstyle[twocolumn,aps]{revtex}
\begin{document}
\draft

\title{Non-trivial aspects of the onset of nuclear collectivity: 
Static moments}

\author{L. Zamick}
\address{Department of Physics and Astronomy, Rutgers University,
	Piscataway, New Jersey 08855}

\author{D. C. Zheng}
\address{W.~K.~Kellogg Radiation Laboratory,
   106-38, California Institute of Technology, Pasadena, California 91125}

\date{\today}

\maketitle

\begin{abstract}
We consider several topics concerning static magnetic dipole and 
electric quadrupole moments ($\mu$ and $Q$) 
as signatures of the onset of nuclear
collectivity. Having previously noted that in $^{50}\mbox{Cr}$
there is an abrupt change of sign in $Q$ of yrast states with 
$J^{\pi}=10^+, 12^+$, and $14^+$
relative to lower $J$ states, we discuss whether these states are oblate
or prolate. We next show that configuration mixing 
leads to much larger changes in $Q$ than in $\mu$. We then look for
other bands of interest in $^{50}\mbox{Cr}$. Finally we 
discuss the Jolos-von Brentano relationship which
relates $Q$ of $2^+_1$ states to $B({\rm E2})$'s for transitions from and 
to the $2^+_1$ states.
\end{abstract}

\vspace{0.3in}

% \pacs{To be found out}

\section{Comment on Static Quadrupole Moments in $jj^{50}\mbox{Cr}$: oblate or
prolate}

In a recent publication \cite{cr50}, 
the current authors noted that in shell model
calculations for $^{50}\mbox{Cr}$, in which up to three nucleons were allowed
to be excited from the $f_{7/2}$ shell to the rest of the $f$-$p$ shell, 
the static quadrupole moments of the yrast states with $J^{\pi}=2^+,
4^+, 6^+$, and $8^+$ were negative but those for 
$J^{\pi}=10^+, 12^+$, and $14^+$ were positive. However, the question
of whether the latter three states were oblate or prolate was not 
answered definitively. We here address this issue.

With the FPD6 interaction \cite{fpd6}, and allowing up to three nucleons 
to be excited from the $f_{7/2}$ shell ($t$=3), the static quadrupole 
moments (in $e\, {\rm fm}^2$) were -27.5, -34.8, -8.1, -20.7 for 
$J^{\pi}_n=2^+_1, 4^+_1, 6^+_1$, and $8^+_1$ 
respectively and were +45.7, +18.6, and
+11.4 for $J^{\pi}_n=10^+_1, 12^+_1$, and $14^+_1$. 
With the KB3 interaction \cite{kb3},
the corresponding values are -24.8, -30.0, -15.6, -14.7 for 
$J^{\pi}_n=2^+_1, 4^+_1, 6^+_1$, and $8^+_1$ and +26.5, +13.0, and +8.2 
for $J^{\pi}_n=10^+_1, 12^+_1$, and $14^+_1$. 
Note that there is not a smooth 
transition in going from $J^{\pi}_n=8^+_1$ to $J^{\pi}_n=10^+_1$. 
The value of $Q$ for $8^+_1$ is fairly large and negative while the value for 
$10^+_1$ is large and positive. 

If $K$ were a good quantum number, we could use the rotational formula 
\begin{equation}
Q(J) = \frac{3K^2-J(J+1)}{(J+1)(2J+3)} Q_K ,   \label{qk}
\end{equation}
where $Q_K$ is the intrinsic quadrupole moment, 
to determine $K$. For $J^{\pi}=10^+, 12^+$, and $14^+$,
if $K$ is small ($K\leq 6$), then $Q(J)$ and $Q_K$ 
have opposite signs. But if $K$ is sufficiently large, 
$Q(J)$ and $Q_K$ will have the same sign. 

We expect considerable band mixing. Nevertheless we feel that a crude 
analysis using the above formula would be helpful in determining 
in which ball park we are. To reduce the ambiguity of the effective
charges, we take ratios. Thus,
\begin{equation}
\frac{Q(10^+)}{Q(12^+)} = 1.387 \frac{3K^2-110}{3K^2-156},
\end{equation}
\begin{equation}
\frac{Q(12^+)}{Q(14^+)} = 1.325 \frac{3K^2-156}{3K^2-210}.
\end{equation}
For the FPD6 interaction \cite{fpd6}, the first equation in the above
gives $K=8.5$ while the second one gives $K=12.1$. 
With the KB3 interaction \cite{kb3},
the corresponding numbers are similar:
$K=9.2$ and $K=12.8$. Thus this admittedly crude analysis favors a
``high-$K$ prolate'' interpretation for these states.

It should be noted that the lower spin states, especially, $J^{\pi}=2^+$
and $4^+$ are best described as low-$K$ prolate states. Thus all 
the states are prolate but the nature of the 
$J^{\pi}=10^+$, $12^+$ and $14^+$ ``band'' is quite different
from that of $J^{\pi}=2^+$, $4^+$, $6^+$ and $8^+$. We clearly have 
a band crossing phenomenon and it is interesting to note that one 
shell model configuration $(f_{7/2})^{10}$ contains in some sense both of 
the two bands.

\section{Magnetic $g$ factors in $^{50}\mbox{Cr}$ and the onset 
of nuclear collectivity}

In a recent experimental work, Pakou {\it et al.} \cite{Pakou2} 
measured $g$ ($g=\mu/J$) factors of states in $^{50}\mbox{Cr}$ with the
following results:
\begin{center}
\begin{tabular}{cc}
$J^{\pi}_n$    &    $g$    \\
$2^+_1$        &   0.54(11)\\
$4^+_1$        &   0.43(9)\\
$6^+_1$        &   0.54(16)\\
$8^+_1$        &   0.54(9)
\end{tabular}
\end{center}

For $4^+_1$, $6^+_1$, and $8^+_1$, these are much smaller than the $g$ 
factors calculated in the single $j$ shell model \cite{Pakou}.
The suggestion was made in the 1994 paper \cite{Pakou2}
that the onset of nuclear
collectivity brought the $g$ factors close to the rotational result 
for a $K=0$ band of $g\simeq g_R = Z/A$. 

This result 
has motivated us here to calculate the $g$ factors in larger shell
model spaces. We allow up to $t$ nucleons to be excited from the $f_{7/2}$
shell to the rest of the $f$-$p$ shell and show results for 
$t=0, 1, 2$, and 3 for the $g$ factors in Table I. 

We should first remark that from our previous work on static quadrupole
moments $Q$ \cite{cr50}, we agree that there is an onset of nuclear 
collectivity, in the sense that the $B({\rm E2})$'s 
become bigger as $t$ increases, the energy levels look more rotational and
$Q$ for $J$ up to 8 become more negative 
relative to $t=0$. In that work, the FPD6 interaction \cite{fpd6}
was used. In this work, we show the behavior of $Q$ 
using the KB3 interaction \cite{kb3}. This also shows 
the increase in magnitude of $Q$ for $J^{\pi}=2^+,4^+,6^+$, 
and $8^+$ (more negative). 

However, when we look at the $g$ factors, the change is not so drastic.
Even for $t=3$ one still gets large $g$ factors. 
The values for $2^+_1$, $4^+_1$, $6^+_1$, and $8^+_1$
using free $g_{l,\pi}$, $g_{l,\nu}$, $g_{s,\pi}$, and $g_{s,\nu}$ 
values are 0.58, 0.80, 0.79, and 0.83, respectively. 
These are considerably larger than the experimental values.

If we use quenched spin $g$ factors $g_{s,\pi/\nu}=0.7 g_{s,\pi/\nu}$, 
along with $g_{l,\pi}= 1.1$ and $g_{l,\nu} = -0.1$, the 
corresponding results for $g$ 
decrease somewhat to 0.54, 0.76, 0.74, and 0.78. 
But they are still substantially larger than experiment.

Thus the {\em calculated} onset of nuclear collectivity consists of 
large changes in the $B({\rm E2})$'s and $Q$, 
but much smaller changes in magnetic $g$ factors. 
With the bare $g_l$ and $g_s$ values, 
the percent change for the $g$ factors in going from $t$=0 to
$t$=3 for $2^+_1$, $4^+_1$, $6^+_1$, $8^+_1$ are 
18.1\%, 15.3\%, 10.4\%, and 7.6\%, respectively. 
As can be seen from Table II, there are more than a factor of two
changes for $Q$. 

It would be nice in the near future to bring about a reconciliation 
between theory and experiment. 

\section{Other ``bands'' in $^{50}\mbox{Cr}$}
In our previous work \cite{cr50}, we focused on yrast states in 
$^{50}\mbox{Cr}$ and showed that whereas the $2^+_1$, $4^+_1$, $6^+_1$, 
and $8^+_1$ states have negative static quadrupole moments $Q$, the 
$10^+_1$, $12^+_1$, and $14^+_1$ have positive $Q$'s. There is a band
crossing and, to some extent, even the simplest configuration 
$(f_{7/2})^{10}$ has in it both the ground-state band and the second 
band which overtakes the ground-state band at $J^{\pi}=10^+$.

In Table III we showed for $t=3$ a common feature of the states
$2^+_2$, $4^+_2$, $6^+_2$, and $8^+_2$. They have rather large, 
positive quadrupole moments. This result contradicts with the 
yrast band calculation for 
which the $Q$'s are comparable in magnitude but are negative. 

We also show in Table III the values of $Q$ for the 
$10^+_1$, $10^+_2$, $12^+_1$, and $14^+_1$ states. They are 
also positive. It is not clear how to extend the band $8^+_2$
-- whether to include the $10^+_1$ or $10^+_2$ state.
Since the two $10^+$ states are rather close in energy, it could 
be that some admixture of these looks most like a member of the 
band.

There have been measurements in other parts of the periodic table
where the $g$ factors for even-even nuclei differ substantially from 
$Z/A$. For example, for $^{150}\mbox{Sm}$, Vass {\it et al.} \cite{koller}
reported that $g(4+)/g(2+)$=1.60(12) whilst $g(6+)/g(2+)$=1.14(34). 
Of course, since in this calculation we are dealing
with $^{50}\mbox{Cr}$ we cannot say that their measurement 
supports our calculation or vice versa.
But at least it suggests that one should be on the lookout
for the types of behaviours that both works seem to find.

\section{Jolos-von Brentano relationship}

Recently R.V.~Jolos and P.~von Brentano 
(hereinafter referred to as J-vB) \cite{jvb}
have presented a formula which relates
quadrupole moments of the $2^+_1$ states to various 
$B({\rm E2})$ values. This connection 
is of great interest because it is much more 
difficult to measure static quadrupole moments than it is to measure 
$B({\rm E2})$'s. They feel that the formula should be extremely accurate 
(better than 1.5\%) for deformed nuclei for which 
$E^*(4^+_1)/E^*(2^+_1) \geq 2.9$, where 
$E^*(4^+_1)$ and $E^*(2^+_1)$ are the excitation energies of the 
$4^+_1$ and $2^+_1$ states relative to the ground state. Also for 
``realistic cases'' the predictions given by the 
formula agree with IBM-1 results to better than $2\%$ for $N=12$ and 
$6\%$ for $N=6$. Their relationship can be written as
\begin{equation}
\frac{|Q(2^+_1)|}{\sqrt{B({\rm E2}: 2^+_1 \rightarrow 0^+_1)}}
= \frac{8}{7} \sqrt{ \pi G(1+R_1-W)},     \label{jvb}
\end{equation}
where
\begin{equation}
G = \left( \frac{7}{10} \right) \frac{B({\rm E2}: 4^+_1 \rightarrow 2^+_1)}
		         {B({\rm E2}: 2^+_1 \rightarrow 0^+_1)},
\end{equation}
\begin{equation}
R_1 = \frac{B({\rm E2}: 2^+_2 \rightarrow 0^+_1)}
	   {B({\rm E2}: 2^+_1 \rightarrow 0^+_1)},
\end{equation}
and
\begin{equation}
W = \frac{B({\rm E2}: 2^+_2 \rightarrow 2^+_1)}
	 {B({\rm E2}: 4^+_1 \rightarrow 2^+_1)}.
\end{equation}

Of course the rotational formulae of Bohr and Mottelson can also 
be combined to give a relationship between $B({\rm E2})$ and $Q(2^+_1)$. 
These are Eq.(\ref{qk}) and 
\begin{equation}
B({\rm E2}: KJ_1 \rightarrow K J_2) 
= \frac{5}{16\pi} e^2 Q_0^2 \; \langle J_1 K 2 0 | J_2 K\rangle^2,
\end{equation}
where $\langle J_1 K 2 0 | J_2 K\rangle$ is the Clebsch-Gordon coefficient.
For a $K=0$ band, one gets
\begin{equation}
|Q(2^+_1)|= \sqrt{\frac{64\pi}{49} B({\rm E2}: 2^+_1\rightarrow 0^+_1)} .
					\label{rot}
\end{equation}
Note that in the rotational limit, $G=1$. If one also takes
$R_1=W=0$, i.e., if one neglects interband transitions, 
one then recovers the 
above rotational formula from the J-vB equation (\ref{jvb}). 
It is interesting to find out if, for a non-perfect rotor, 
the J-vB relation would yield a more accurate $Q(2^+_1)$. 
To this end, we conduct a theoretical experiment by performing 
shell-model calculations for the $B({\rm E2})$ values that
go into Eqs.~(\ref{jvb}) and (\ref{rot}) and comparing the predictions
of these two formulae for $|Q(2^+_1)|$ to the 
``exact'' values obtained in the shell-model calculations. 
We do this calculation for selected deformed nuclei 
in the $s$-$d$ ($^{20}\mbox{Ne}$, $^{22}\mbox{Ne}$, $^{24}\mbox{Mg}$ and
$^{28}\mbox{Si}$) and $f$-$p$ 
($^{46}\mbox{Ti}$, $^{48}\mbox{Ti}$, and $^{50}\mbox{Cr}$) region.
For the $s$-$d$ shell, we use the Brown-Wildenthal interaction 
\cite{BW}; for the $f$-$p$ shell, we use the modified Kuo-Brown 
interaction KB3 \cite{kb3}. For the $s$-$d$ shell nuclei, the calculations 
are carried out in the full one-major-shell space. For the $f$-$p$ shell nuclei,
the full space calculation is only done for $^{46}\mbox{Ti}$. For the other
two nuclei, a maximum number of $t$ nucleons are allowed to leave the 
$f_{7/2}$ orbital and occupy the rest of the $f$-$p$ shell with 
$t=4$ for $^{48}\mbox{Ti}$ and 
$t=3$ for $^{50}\mbox{Cr}$.

Our results are listed in Table IV where we also list the experimental
values for the ratio $E^*(4^+_1)/E^*(2^+_1)$ and $Q(2^+_1)$.
The calculated values for various $B({\rm E2})$'s
that go into the J-vB formula (\ref{jvb}) and the rotational formula
(\ref{rot}) are listed in Table V where
the experimental $B({\rm E2}: 2^+_1\rightarrow 0^+_1)$
values are also shown. 
With one notable exception, the J-vB predictions 
agree with the shell model results to better than 12\%. 
However, for all the nuclei that are considered here, 
only for one nucleus ($^{46}\mbox{Ti}$) has the J-vB formula done a 
better job than the rotational formula.
This is surprising because one would expect
that there is more physics put into the J-vB formula.

The biggest disagreement between the rotational formula and the 
shell model results occurs in $^{46}\mbox{Ti}$ and $^{48}\mbox{Ti}$, 
where the discrepancies are 34\% and 37\% respectively. 
The J-vB formula seems to cure the problem for $^{46}\mbox{Ti}$
but not for $^{48}\mbox{Ti}$. The problem in the latter case is that $R_1$ 
and $W$ are almost the same and so cancel each other out.

In Table VI we apply the J-vB relation to experimental input in the 
$f$-$p$ shell which are obtained from the Nuclear Data Sheets \cite{nds}. 
Note that the experimental $B({\rm E2})$'s are somewhat larger 
than those calculated with the KB3 interaction with effective charges
of $e_p$=1.5 and $e_n$=0.5. The experimental values of $G$ are considerably
smaller than the calculated values. In other words, 
$B({\rm E2}: 4^+_1 \rightarrow 2^+_1)_{\rm exp}$ is smaller than 
$B({\rm E2}: 4^+_1 \rightarrow 2^+_1)_{\rm theory}$. There are considerable
differences in the values of $R_1$ and $W$ as well. 

Using the J-vB relation with experimental data the values of 
$Q(2^+_1)$ are significantly smaller than those using the rotational
model. For $^{46}\mbox{Ti}$, $^{48}\mbox{Ti}$, $^{50}\mbox{Cr}$,
the J-vB (rotational) values of $Q(2^+_1)$ are respectively 
20.5 (28.7), 12.7 (24.3), and 22.5 (29.8). For 
$^{46}\mbox{Ti}$ and $^{48}\mbox{Ti}$, the J-vB analysis gives an 
improved fit. For $^{50}\mbox{Cr}$, the J-vB analysis gives too small 
a value of $Q(2^+)$ compared with experiment. 

It is difficult to give a definite assessment of the J-vB relation 
in the regions that we have considered (which in some cases are beyond
what the authors envisioned). One problem is that the error bars
on the data are rather large and they may be systematic errors beyond 
those taken into account. But it would seem that the J-vB relation 
works better when experimental data is used rather than 
``theoretical data'' from shell model calculations.

\acknowledgements
This work was supported in part by a Department of Energy grant
DE-FG05-86ER-40299 (L.Z.). We thank Noemie Koller for her interest and
help.

% \pagebreak

\begin{table}

TABLE I. $g$ factors ($g=\mu/J$) in $^{50}\mbox{Cr}$ for 
the KB3 interaction as a function of $t$, the maximum number 
of nucleons allowed to be excited from the $f_{7/2}$ shell to the 
rest of the $f$-$p$ shell.

\begin{tabular}{cccccc}
$J$ & $t$=0$^{a)}$ & $t$=1$^{a)}$ & $t$=2$^{a)}$ & $t$=3$^{a)}$ & $t$=3$^{b)}$
					\\ \hline
2 & 0.707 & 0.679 & 0.578 & 0.579 & 0.540 \\
4 & 0.949 & 0.894 & 0.831 & 0.804 & 0.756 \\
6 & 0.885 & 0.858 & 0.816 & 0.792 & 0.745 \\
8 & 0.769 & 0.822 & 0.841 & 0.828 & 0.779 \\
10& 0.486 & 0.519 & 0.515 & 0.509 & 0.474 \\
12& 0.609 & 0.609 & 0.591 & 0.588 & 0.550 \\
14& 0.712 & 0.698 & 0.687 & 0.678 & 0.633
\end{tabular}

$^{a)}$ For free $g$ values:
	$g_{l,\pi}$=1, $g_{l,\nu}$=0, $g_{s,\pi}$=5.586, $g_{s,\nu}$=-3.826.

$^{b)}$ For renormalized $g$ values:
	$g_{l,\pi}$=1.1, $g_{l,\nu}$=-0.1, 
        $g_{s,\pi}$=3.910, $g_{s,\nu}$=-2.678.

\end{table}

\begin{table}

TABLE II. Static quadrupole moments $Q$ (in units of $e\, {\rm fm}^2$)
in $^{50}\mbox{Cr}$ for the KB3 interaction as a function of $t$, 
the maximum number 
of nucleons allowed to be excited from the $f_{7/2}$ shell to the 
rest of the $f$-$p$ shell.

\begin{tabular}{ccccc}
$J$ & $t$=0 & $t$=1 & $t$=2 & $t$=3 \\ \hline
2 & -12.240 & -20.392 & -20.824 & -24.665 \\
4 & -12.148 & -22.792 & -24.950 & -29.810 \\
6 &  -4.415 & -14.459 &  -9.661 & -15.531 \\
8 &   0.478 &  -8.490 & -10.454 & -14.698 \\
10&  19.118 &  23.481 &  24.494 &  26.461 \\
12&   6.546 &  10.488 &  11.591 &  12.998 \\
14&   6.810 &   8.759 &   8.208 &   8.232 
\end{tabular}

\end{table}

\begin{table}

TABLE III. Other possible positive-parity bands in $^{50}\mbox{Cr}$ 
in the $t$=3 calculation with the KB3 interaction.

\begin{tabular}{cccc}
$J^{\pi}_n$ & $E_x$(MeV) & $\mu (\mu_N)$ & $Q (e\, {\rm fm}^2)$ \\ \hline
$ 4^+_2$ & 3.003 & 5.680 & 31.853 \\
$ 6^+_2$ & 3.595 &-0.414 & 40.262 \\
$ 8^+_2$ & 5.611 & 2.172 & 19.469 \\
$10^+_1$ & 5.993 & 5.095 & 26.461 \\
$10^+_2$ & 6.500 & 6.137 & 12.468 \\
$12^+_1$ & 7.435 & 7.058 & 12.998 \\
$14^+_1$ & 9.949 & 9.490 &  8.232

\end{tabular}

\end{table}

\begin{onecolumn}

\begin{table}

TABLE IV. The experimental (exp) values \cite{spear,towsley,lesser}
and the results of shell model (SM), 
Jolos-von Brentano (J-vB) and rotational (rot) formulae 
for static quadrupole moments [$Q(2^+_1)$ for ``exp'' and ``SM''; 
$|Q(2^+_1)|$ for ``J-vB'' and ``rot'']
(in $e\, {\rm fm}^2$) of $2^+_1$ states in selected $s$-$d$ and $f$-$p$ shell
nuclei. The predictions of the J-vB and rotational formulae based 
on the shell model $B({\rm E2})$ values should be
compared with the shell model results. 
In the parentheses we give the percentage deviations of the ``J-vB'' and 
``rot'' results from the shell model. Effective charges
$e_p=1.5$ and $e_n=0.5$ are assumed. 
\begin{tabular}{ccccccc}
Nucleus & $\frac{E^*(4^+_1)}{E^*(2^+_1)}_{\rm exp}$ 
        & $\frac{E^*(4^+_1)}{E^*(2^+_1)}_{\rm SM}$ 
 & $Q(2^+_1)_{\rm exp}$ & $Q(2^+_1)_{\rm SM}$  
 & $|Q(2^+_1)_{\rm J\!-\!vB}|$ & $|Q(2^+_1)_{\rm rot}|$  \\ \hline
$^{20}\mbox{Ne}$&2.61 & 2.37 & $-23\pm 3$
		&-15.83 & 13.96(-11.8\%) & 15.78(-0.3\%)\\
$^{22}\mbox{Ne}$&2.65 & 2.47 & $-19\pm 4$
		&-15.67 & 15.92(+1.6\%) & 15.89(+1.4\%)\\
$^{24}\mbox{Mg}$&3.01 & 2.90 &$-18\pm 2$
		&-19.25 & 18.46(-4.1\%) &19.90(+3.3\%)\\
$^{28}\mbox{Si}$&2.60 & 2.34 &$-16\pm 3$
		& 20.75 & 19.18(-7.6\%) & 20.25(-2.4\%)\\
$^{46}\mbox{Ti}$&2.26 & 1.90 &$-21\pm 6$
		&-17.30 & 17.72(+2.4\%) & 23.21(+34.2\%)\\
$^{48}\mbox{Ti}$&2.33 & 2.25 &$-13.5\pm 8.8$
		&-14.72 & 20.20(+37.2\%) & 20.17(+37.0\%)\\
$^{50}\mbox{Cr}$&2.40 & 2.35 &$-36\pm 7$
		&-24.82 & 27.31(+10.0\%) & 26.63(+7.3\%)
\end{tabular}

\end{table}

\begin{table}

TABLE V. Input from shell model calculations into the 
J-vB and rotational formulae, obtained for the Wildenthal interaction
for the $s$-$d$ shell and the KB3 interaction for the $f$-$p$ shell.
The $B({\rm {\rm E2}})$ values listed are in units of $e^2\, {\rm fm}^4$. 
The ratios $G$, $R_1$, and $W$ are defined in the text. 
We also give the experimental $B({\rm E2}: 2^+_1 \rightarrow 0^+_1)$ 
values (in the parentheses). 

\begin{tabular}{ccccccccc}
Nucleus & $B({\rm E2}: 2^+_1 \rightarrow 0^+_1)$  & (exp)
        & $B({\rm E2}: 2^+_2 \rightarrow 0^+_1)$ 
	& $B({\rm E2}: 2^+_2 \rightarrow 2^+_1)$ 
	& $B({\rm E2}: 4^+_1 \rightarrow 2^+_1)$ 
	& $G$ & $R_1$ & $W$ \\ \hline
$^{20}\mbox{Ne}$ & 60.67 &(68)   & 0.03 & 4.41 & 72.20 &0.83 &0.001 &0.061 \\
$^{22}\mbox{Ne}$ & 61.53 &(46)   & 4.57 & 0.55 & 82.62 &0.94 &0.074 &0.0066 \\
$^{24}\mbox{Mg}$ & 96.48 &(86.4) & 8.63 &21.07 &128.23 &0.93 &0.089 &0.164 \\
$^{28}\mbox{Si}$ & 99.92 &(65.2) & 0.36 &13.86 &141.41 &0.99 &0.004 &0.098 \\
$^{46}\mbox{Ti}$ &131.32 &(201)  & 4.48 &70.25 &173.59 &0.93 &0.034 &0.405 \\
$^{48}\mbox{Ti}$ & 99.10 &(144)  &23.44 &40.99 &148.02 &1.05 &0.236 &0.277 \\
$^{50}\mbox{Cr}$ &172.82 &(216)  &11.27 & 2.13 &245.72 &1.00 &0.065 &0.0087
\end{tabular}

\end{table}

\begin{table}

TABLE VI. Same as Table V but using experimental input. Only the 
results for  the $f$-$p$ shell nuclei are listed. 

\begin{tabular}{ccccccc}
Nucleus & $B({\rm E2}: 2^+_1 \rightarrow 0^+_1)$ 
   & $G$ & $R_1$ & $W$ & $|Q(2^+_1)_{\rm J\!-\!vB}|$
		       & $|Q(2^+_1)_{\rm rot}|$ \\ \hline
$^{46}\mbox{Ti}$ & 201   &0.686 & 0.003 & 0.260 & 20.5  & 28.7 \\
$^{48}\mbox{Ti}$ & 144   &0.564 & 0.060 & 0.580 & 12.6  & 24.3 \\
$^{50}\mbox{Cr}$ & 216   &0.516 & 0.106 & 0.000 & 22.5$^{a)}$  & 29.8 \\
                 &       &      &       & 0.219 & 20.1$^{b)}$  & 29.8 \\
\end{tabular}

$^{a)}$ We consider only $2^+_2 \rightarrow 0^+_1$ transition to determine $W$.

$^{b)}$ We add $2^+_3 \rightarrow 0^+_1$ and $2^+_3 \rightarrow 0^+_1$ 
        transitions. The states are close together:
	$E(2^+_2)=2.924 {\rm MeV}$ and $E(2^+_3)=3.161 {\rm MeV}$.

\end{table}

\end{onecolumn}


\begin{references}
\bibitem{cr50} L.~Zamick, M.~Fayache, and D.C.~Zheng, Phys. Rev. C {\bf 53},
	188 (1996). 
\bibitem{fpd6} W.A.~Richter, M.G.~Van Der Merwe, R.E.~Julies, and 
	B.A.~Brown, Nucl. Phys. {\bf A523}, 325 (1991).
\bibitem{kb3} A.~Poves and A.P.~Zuker, Phys. Rep. {\bf 70}, 235 (1981).
\bibitem{Pakou2} A.~Pakou, J.B.~Billows, A.W.~Monteford, and 
	D.D.~Warner, Phys. Rev. C {\bf 50}, 2608 (1994).
\bibitem{Pakou} A.~Pakou, R.~Tanczyn, D.~Turner, W.~Jan, G.~Kumbartzki, 
	N.~Benczer-Koller, Xiao-li Li, Huan Liu, 
	and L.~Zamick, Phys. Rev. C {\bf 36}, 2088 (1987).
\bibitem{koller} T.~Vass, A.W.~Mountford, G.~Kumbartzki, 
	N.~Benczer-Koller and R.~Tanczyn, Phys. Rev. C {\bf 48},2640 (1993).
\bibitem{jvb} R.V.~Jolos and P.~von Brentano, preprint.
\bibitem{BW} B.A.~Brown and B.H.~Wildenthal, Ann. Rev. Nucl. Part. Sci. 
	{\bf 38}, 29 (1988).
\bibitem{spear} R.H.~Spear, Physics Reports {\bf C 73}, 369 (1981).
\bibitem{towsley} C.W.~Towsley, D.~Cline, and R.N.~Horoshko, 
	Nucl. Phys. {\bf A 250}, 381 (1975).
\bibitem{lesser} R.M.S.~Lesser, D.~Cline, P.~Goode, and 
	R.N.~Horoshko, Nucl. Phys. {\bf A 190}, 579 (1972).
\bibitem{nds} Nuclear Data Sheets {\bf 68}, 1 (1993); 
	{\it loc. cit.,} {\bf 68}, 271 (1993);
	{\it loc. cit.,} {\it loc. cit.,} {\bf 75}, 1 (1995).
\end{references}
\end{document}